\documentclass[reprint,superscriptaddress,showpacs,amsmath,amssymb,aps,pra]{revtex4-1}
\usepackage{times}
\usepackage{amssymb}
\usepackage{graphicx}
\usepackage{dcolumn}
\usepackage{dsfont}
\usepackage{bm}
\usepackage{subfigure}
\usepackage[ruled]{algorithm2e}
\usepackage{hyperref}
\usepackage{appendix}

\hypersetup{colorlinks=true, citecolor=blue, urlcolor=blue, linkcolor=blue}
\begin{document}
\title{Quantum algorithms for the generalized eigenvalue problem}
\author{Jin-Min Liang}
\email{jmliang@cnu.edu.cn.}
\affiliation{School of Mathematical Sciences, Capital Normal University, Beijing 100048, China}
\author{Shu-Qian Shen}
\affiliation{College of Science, China University of Petroleum, Qingdao 266580, China}
\author{Ming Li}
\affiliation{College of Science, China University of Petroleum, Qingdao 266580, China}
\author{Shao-Ming Fei}
\email{feishm@cnu.edu.cn.}
\affiliation{School of Mathematical Sciences, Capital Normal University, Beijing 100048, China}
\affiliation{Shenzhen Institute for Quantum Science and Engineering, Southern University of Science and Technology, Shenzhen 518055, China}
\date{\today}

\begin{abstract}
The generalized eigenvalue (GE) problems are of particular importance in various areas of science engineering and machine learning. We present a variational quantum algorithm for finding the desired generalized eigenvalue of the GE problem, $\mathcal{A}|\psi\rangle=\lambda\mathcal{B}|\psi\rangle$, by choosing suitable loss functions. Our approach imposes the superposition of the trial state and the obtained eigenvectors with respect to the weighting matrix $\mathcal{B}$ on the Rayleigh quotient. Furthermore, both the values and derivatives of the loss functions can be calculated on near-term quantum devices with shallow quantum circuit. Finally, we propose a full quantum generalized eigensolver (FQGE) to calculate the minimal generalized eigenvalue with quantum gradient descent algorithm. As a demonstration of the principle, we numerically implement our algorithms to conduct a 2-qubit simulation and successfully find the generalized eigenvalues of the matrix pencil $(\mathcal{A},\,\mathcal{B})$. The numerically experimental result indicates that FQGE is robust under Gaussian noise.
\end{abstract}

\maketitle

\section{Introduction}
It has been demonstrated that quantum computer can solve problems efficiently that are intractable on classical computer. Some powerful quantum algorithms have been presented such as factoring \cite{Shor1994}, database searching \cite{Grover1996}, matrix inverse \cite{HHL2009}. These algorithms have recently
been employed in machine learning such as regression and classification problems \cite{QML2017,liu2018quantum,liang2019quantum}. However, these algorithms require long coherence time and high-fidelity gates \cite{preskill2018quantum}. Before the emergence of large-scale, fault-tolerate universal quantum computer, the noisy intermediate-scale quantum (NISQ) processors are thought to be a significant intermediate product. A kind of hybrid quantum-classical algorithms, so-called variational quantum algorithms (VQAs), plays a crucial role on NISQ era \cite{preskill2018quantum}. VQAs aim to tackle complex problems by using classical computer and NISQ devices, such as finding the energy spectra \cite{peruzzo2014variational,higgott2019variational,jones2019variational,vogt2021preparing}, simulating the Schr$\ddot{\textrm{o}}$dinger equations \cite{li2017,mahdian2020incoherent,endo2020variational} and quantum machine learning \cite{benedetti2019parameterized,wang2020variational,li2021optimizing}. The classical computer finds the optimal parameter by optimizing the loss function designed for the problem. The value of loss function and its gradient information are calculated entirely on the NISQ devices by performing measurements. Compared with perfect quantum computers, this hybridization reduces the quantum resources including the numbers of gates and qubits, the circuit depth and the numbers of measurements \cite{larose2019variational}.

The solutions of Schr$\ddot{\textrm{o}}$dinger equations are often determined by standard large-scale eigenvalue problems in quantum chemistry and many-body quantum systems. Many approaches have been used to tackle such standard eigenvalue problems. In particular, given a Hamiltonian $\mathcal{H}$ of a quantum system, one performs an classical optimization procedure on the loss function $C=\langle\psi|\mathcal{H}|\psi\rangle$ with a trial state $|\psi\rangle$. The minimal value of the loss function $C$ implies the ground energy and its corresponding state is thought to be an approximate lowest eigenvector \cite{peruzzo2014variational}. This hybrid quantum-classical algorithm is the well-known variational quantum eigensolver (VQE). Once the ground state has been obtained, one can define an updated Hamiltonian which ground state is the first excited state of original Hamiltonian \cite{jones2019variational}. Reference \cite{higgott2019variational} finds the $k$-th excited state of a Hamiltonian $\mathcal{H}$ by a variational quantum deflation method. Their approach adds the overlap term between eigenstates onto the original loss function. Another way for discovering excited states is subspace VQE which contains weighted and non-weighted forms \cite{nakanishi2019subspace}. Subspace expansion approach \cite{mcclean2017hybrid} and multistate contracted VQE \cite{parrish2019quantum} approximate each eigenstate as a linear combination of a set of orthogonal states. The training coefficients are obtained by solving a GE problem. Different form the standard VQE \cite{peruzzo2014variational}, the work \cite{wei2020a} described a full quantum eigensolver for quantum chemistry simulations. The key idea of \cite{wei2020a} is a quantum version of gradient descent algorithm which can be implemented totally on a NISQ computer.

However, the Schr$\ddot{\textrm{o}}$dinger equation for hydrogen and helium atoms is written as a generalized eigenvalue (GE) problem \cite{amos1969a} which differs from the standard eigenvalue problem. Thus, a natural question arise that how to solve the GE problem with quantum technology. Large-scale GE problems arise in various of science and engineering, such as linear stability analysis of the Navier-Stokes equation in fluid dynamics \cite{cliffe2000the}, the vibration of a cantilever beam \cite{bittnar1996numerical,ghaboussi2016numerical} and the electron energy and position problems in quantum chemistry \cite{ford1974the}. In many specific examples, the non-self-adjoint eigenvalue problem of infinite-dimensional Hamiltonian systems can also be written as the GE problem \cite{gantmacher2002oscillation,chugunova2010count}. Although several powerful tools have been developed \cite{golub1989matriz,ericsson1980the,sakuraia2003a,ikegamia2009a}, the computation of GE problems is still a challenge for classical computer. More recently, Parker and Joseph \cite{parker2020quantum} presented an approach with the help of quantum phase estimation (QPE). However, as pointed in \cite{peruzzo2014variational}, the QPE requires long coherence time and makes this approach unsuitable for the NISQ devices.

In this work, we introduce two types of quantum algorithms for GE problems. For the first type, a variational quantum generalized eigensolver (VQGE) for finding the desired generalized eigenvalues of a GE problem, $\mathcal{A}|\psi\rangle=\lambda\mathcal{B}|\psi\rangle$, is presented by utilizing the powerful NISQ technology, where $\mathcal{A}\in\mathbb{R}^{N\times N}$ is an Hermitian matrix and $\mathcal{B}\in\mathbb{R}^{N\times N}$ is a positive definite Hermitian matrix. We design different loss functions to find different generalized eigenvalues. For the second type, inspired by the full quantum eigensolver \cite{wei2020a}, we adapt the quantum gradient descent algorithm to find a lowest generalized eigenvalue. We call this generalized eigensolver a full quantum generalized eigensolver (FQGE). Finally, we numerically simulate our algorithms to solve a 2-qubit GE problem on quantum cloud platform and successfully obtain the desired result.

\section{Variational method: a variational quantum generalized eigensolver}

\subsection{Theoretical basis of VQGE}

Let $\mathcal{A}\in\mathbb{R}^{N\times N}$ be an Hermitian matrix and $\mathcal{B}\in\mathbb{R}^{N\times N}$ a positive definite Hermitian matrix. The GE problem is defined by
\begin{equation}\label{GEP}
\mathcal{A}|\psi\rangle=\lambda\mathcal{B}|\psi\rangle,
\end{equation}
where $|\psi\rangle$ is an eigenvector with generalized eigenvalue $\lambda$ \cite{golub1989matriz}. A significant assumption here is that the matrices $\mathcal{A}$ and $\mathcal{B}$ have decompositions:
\begin{equation}\label{decomposition}
\begin{aligned}
&\mathcal{A}=\sum_{k=0}^{K-1}\alpha_k\mathcal{A}_{k},\quad\mathcal{B}=\sum_{l=0}^{L-1}\beta_l\mathcal{B}_{l},
\end{aligned}
\end{equation}
where $\mathcal{A}_{k}$, $\mathcal{B}_{l}$ are unitaries which can be easily implemented on an NISQ computer \cite{liu2020variational}. We also assume that the number of the terms $L$ and $K$ scale polynomially with the number of qubits, $\mathcal{O}(poly\log N)$, and there are $r$ different generalized eigenvalues ordered in increasing order, $\lambda_1<\lambda_2<\cdots<\lambda_{r}$.

In order to estimate the minimal and maximal generalized eigenvalues, we need to implement the classical optimization procedure for the following loss function,
\begin{equation}\label{function1}
\begin{aligned}
\mathcal{F}(\boldsymbol\theta)
=\frac{\textrm{Tr}[\mathcal{A}\mathcal{U}(\boldsymbol\theta)\rho\mathcal{U}^{\dag}(\boldsymbol\theta)]}
{\textrm{Tr}[\mathcal{B}\mathcal{U}(\boldsymbol\theta)\rho\mathcal{U}^{\dag}(\boldsymbol\theta)]},
\end{aligned}
\end{equation}
where $\rho=|\psi_{\textrm{in}}\rangle\langle\psi_{\textrm{in}}|$ is an arbitrary initial $\log_2N$-qubit state. Eq. (\ref{function1}) can be viewed as a Rayleigh quotient of any state $|\psi(\boldsymbol\theta)\rangle$ which is prepared via a parameterized quantum circuit $\mathcal{U}(\boldsymbol\theta)$ acting on an initial state $|\psi_{\textrm{in}}\rangle$ with an adjustable gate parameter $\boldsymbol\theta$. For any state $|\psi(\boldsymbol\theta)\rangle$, we have $\lambda_{1}\leq\mathcal{F}(\boldsymbol\theta)\leq\lambda_{r}$ \cite{parlett1998symmetric}, with
$$
\lambda_1=\min_{\boldsymbol\theta}\mathcal{F}(\boldsymbol\theta),\quad\lambda_r=\max_{\boldsymbol\theta}\mathcal{F}(\boldsymbol\theta),
$$
associated with two optimal parameters $\boldsymbol\theta_1^{*}$ and $\boldsymbol\theta_r^{*}$ such that
$|\psi_1(\boldsymbol\theta_1^{*})\rangle=\mathcal{U}(\boldsymbol\theta_1^{*})|\psi_{\textrm{in}}\rangle$ and
$|\psi_r(\boldsymbol\theta_r^{*})\rangle=\mathcal{U}(\boldsymbol\theta_r^{*})|\psi_{\textrm{in}}\rangle$, respectively.

When the matrix $\mathcal{B}$ is positive definite, the eigenvectors of different generalized eigenvalues are mutually $\mathcal{B}$-orthogonal \cite{golub1989matriz},
\begin{equation}
\begin{aligned}
\langle\psi_i|\mathcal{B}|\psi_j\rangle=\delta_{ij},\quad\mathcal{A}|\psi_j\rangle=\lambda_j\mathcal{B}|\psi_j\rangle,
\end{aligned}
\end{equation}
where $\delta_{ij}$ denotes the Kronecker delta function. Hence, concerning other generalized eigenvalues and the associated eigenvectors, we define the following new loss functions,
\begin{equation}\label{function2}
\begin{aligned}
\mathcal{F}_j(\boldsymbol\theta)=\mathcal{F}(\boldsymbol\theta)
+\sum_{i=1}^{j-1}\gamma_i\textrm{Tr}[\mathcal{B}\mathcal{U}(\boldsymbol\theta)\rho\mathcal{U}^{\dag}(\boldsymbol\theta_i^{*})]^2,
\end{aligned}
\end{equation}
where $|\psi(\boldsymbol\theta_{i}^{*})\rangle$ is the known eigenvector determined previously. The parameters $\gamma_i$ will be determined below so that the minimum of the loss function $\mathcal{F}_j(\boldsymbol\theta)$ is the $j$th eigenvalue. One can think of this loss function (\ref{function2}) as minimizing $\mathcal{F}(\boldsymbol\theta)$ under the constraint $\textrm{Tr}[\mathcal{B}\mathcal{U}(\boldsymbol\theta)\rho\mathcal{U}^{\dag}(\boldsymbol\theta_i^{*})]^2=0$. In this case, the $j^{\textrm{th}}$ generalized eigenvalues $\lambda_j=\min_{\boldsymbol\theta}\mathcal{F}_j(\boldsymbol\theta)$ and associated eigenvector $|\psi_r(\boldsymbol\theta_{j}^{*})\rangle=\mathcal{U}(\boldsymbol\theta_{j}^{*})|\psi_{\textrm{in}}\rangle$.

\subsection{The computation of loss functions}
The functions $\mathcal{F}(\boldsymbol\theta)$ and $\mathcal{F}_j(\boldsymbol\theta)$ are calculated by computing the following terms, $\langle\mathcal{A}\rangle=\textrm{Tr}[\mathcal{A}\mathcal{U}(\boldsymbol\theta)\rho\mathcal{U}^{\dag}(\boldsymbol\theta)]$, $\langle\mathcal{B}\rangle=\textrm{Tr}[\mathcal{B}\mathcal{U}(\boldsymbol\theta)\rho\mathcal{U}^{\dag}(\boldsymbol\theta)]$ and $\textrm{Tr}[\mathcal{B}\mathcal{U}(\boldsymbol\theta)\rho\mathcal{U}^{\dag}(\boldsymbol\theta_i^{*})]$. The first two items can be abstracted to a general form $\langle\mathcal{H}\rangle=\sum_{i=0}^{d-1}a_i\langle\mathcal{H}_i\rangle$ which the real and imaginary parts can be efficiently evaluated via the Hadamard test \cite{aharonov2008a} or the projective measurements \cite{nielsen2000quantum}. If the desired error of $\langle\mathcal{H}\rangle$ is $\epsilon_{\mathcal{H}}$, the error of $\langle\mathcal{H}_i\rangle$ satisfies \cite{romero2018strategies},
$$
\epsilon_i^2=\frac{|a_i|\epsilon_{\mathcal{H}}^2}{\sum_{i=0}^{d-1}|a_i|}.
$$
The number of measurements $M_i$ required to estimate $\langle\mathcal{H}_i\rangle$ is at most $\mathcal{O}(|a_i|^2\epsilon_i^{-2})$ \cite{romero2018strategies}. The last item is a sum of inner products,
\begin{equation}
\begin{aligned}
\textrm{Tr}[\mathcal{B}\mathcal{U}(\boldsymbol\theta)\rho\mathcal{U}^{\dag}(\boldsymbol\theta_i^{*})]^2=
|\langle\psi(\boldsymbol\theta)|\mathcal{B}|\psi(\boldsymbol\theta_{i}^{*})\rangle|^2.
\end{aligned}
\end{equation}
Several existed approaches are developed for estimating the overlap \cite{buhrman2001quantum,carlos2013swap,cincio2018learning}. The origin swap test determines the overlap to precision $\epsilon$ with sample cost $O(\epsilon^{-2})$ and the circuit depth $O(poly\log N)$ \cite{buhrman2001quantum}. In \cite{carlos2013swap,cincio2018learning}, the modified swap test reduces the depth of circuit to $O(1)$ using parallel Bell-basis measurements and classical logic.

Our states, however, are prepared by the structure unitary circuit $\mathcal{U}(\boldsymbol\theta)$. We utilize a more practical method introduced in \cite{havlivcek2019supervised}. If we are given an initial state $|\boldsymbol{0}\rangle=|0^{\otimes\log N}\rangle$, the explicit value is
\begin{align}
|\langle\psi(\boldsymbol\theta)|\mathcal{B}|\psi(\boldsymbol\theta_{i}^{*})\rangle|^2
&=\Big|\langle\boldsymbol{0}|\mathcal{U}^{\dag}(\boldsymbol\theta)
\sum_{l=0}^{L-1}\beta_l\mathcal{B}_l\mathcal{U}(\boldsymbol\theta_{i}^{*})|\boldsymbol{0}\rangle\Big|^2\\
&=\Big|\langle\boldsymbol{0}|\mathcal{U}^{\dag}(\boldsymbol\theta)
\sum_{l=0}^{L-1}\beta_l\mathcal{B}_l\mathcal{U}(\boldsymbol\theta_{i}^{*})|\boldsymbol{0}\Big|^2\\
&=\Big|\sum_{l=0}^{L-1}\beta_l\langle\boldsymbol{0}|\mathcal{U}^{\dag}(\boldsymbol\theta)\mathcal{B}_l
\mathcal{U}(\boldsymbol\theta_{i}^{*})|\boldsymbol{0}\rangle\Big|^2.
\end{align}
In this case, we only perform the circuit $\mathcal{U}^{\dag}(\boldsymbol\theta)\mathcal{B}_l\mathcal{U}(\boldsymbol\theta_{i}^{*})$ on state $|\boldsymbol{0}\rangle$ and then obtain the real and imaginary part $\textrm{Re}_{li},\textrm{Im}_{li}$ of each term $\langle\boldsymbol{0}|\mathcal{U}^{\dag}(\boldsymbol\theta)\mathcal{B}_l
\mathcal{U}(\boldsymbol\theta_{i}^{*})|\boldsymbol{0}\rangle$ via the Hadamard test \cite{aharonov2008a}. As a result, we have
\begin{align}
\textrm{Tr}[\mathcal{B}\mathcal{U}(\boldsymbol\theta)\rho\mathcal{U}^{\dag}(\boldsymbol\theta_i^{*})]^2=
\Bigg[\sum_{l=0}^{L-1}\beta_{l}(\textrm{Re}_{li}+\textrm{Im}_{li})\Bigg]^2.
\end{align}

\subsection{Optimization of the loss functions}
To find the optimal parameters $\{\boldsymbol\theta_{i}^{*}\}$, we apply the Adam, an algorithm for first-order gradient-based optimization of stochastic objective functions \cite{kingma2014adma} in a variational quantum circuit. Methods based on gradient-free are also available to do the optimization. Starting from a random initial parameter $\boldsymbol\theta^{(1)}$, the $s$th iterated parameter is given by, $\boldsymbol\theta^{(s+1)}=\boldsymbol\theta^{(s)}-\delta\nabla\mathcal{F}(\boldsymbol\theta^{(s)})$, where $\delta$ is the learning rate and $\nabla\mathcal{F}(\boldsymbol\theta^{(s)})$ is the gradient of function $\mathcal{F}(\boldsymbol\theta)$ with respect to the parameter $\boldsymbol\theta^{(s)}$. By convergence, the optimal quantum circuit produces an approximate eigenvector $|\psi(\boldsymbol\theta^{*})\rangle=\mathcal{U}(\boldsymbol\theta^{*})|\psi_{\textrm{in}}\rangle$.
The improved technique introduced in \cite{wierichs2020avoiding,dean2020avoiding} can be used to avoid local minimal.

To estimate the gradient of $\mathcal{F}(\boldsymbol\theta)$ we present the variational quantum circuit $\mathcal{U}(\boldsymbol\theta)$, see Fig. 1,
\begin{equation}
\begin{aligned}
\mathcal{U}(\boldsymbol\theta)=\mathcal{U}_{ent}\mathcal{U}_{R}^{L}(\theta_L)\cdots\mathcal{U}_{ent}\mathcal{U}_{R}^{2}(\theta_2)
\mathcal{U}_{ent}\mathcal{U}_{R}^{1}(\theta_1),
\end{aligned}
\end{equation}
constituted of single-qubit rotations and two-qubit entangling gates \cite{HardwareVQE2017,havlivcek2019supervised}, where
$\theta_t=(\theta_{1}^{t},\cdots,\theta_{n}^{t})^{\dag}$, $t=1,\cdots,L$. The entangling gate is given by $\mathcal{U}_{ent}=\Pi_{(i,j)}Z(i,j)$ with a sequence of CNOT gate $Z(i,j)$ applied on the qubit pair $(i,j)$. The single-qubit operation is given by $\mathcal{U}_{R}^t(\theta_t)=\otimes_{i=1}^n\mathcal{U}(\theta_i^t)$ which is a tensor product of local unitary operators $\mathcal{U}(\theta_i^t)\in\textrm{SU}(2)$. As a detailed example of this type of local unitary operators, we consider $\mathcal{U}(\theta_i^t)=R_y(\theta_i^t)=e^{\textrm{i}\frac{\theta_{i}^t}{2}Y}$, where $Y$ is the second Pauli matrix and $\theta_i^t\in[0,2\pi]$. For such a circuit with depth $L$, the total number of parameters is $nL$. Since the number of parameters grow linearly with the number of qubits and the circuit depth, only polynomial measurements are required. The following theorem shows that the gradient of loss functions can be calculated by rotating the parameter $\theta_i^t$ by $\pi$, which can be easily implemented on a near-term computer, see proof in Appendix A.
\begin{figure}[]
\centering
\includegraphics[scale=0.7]{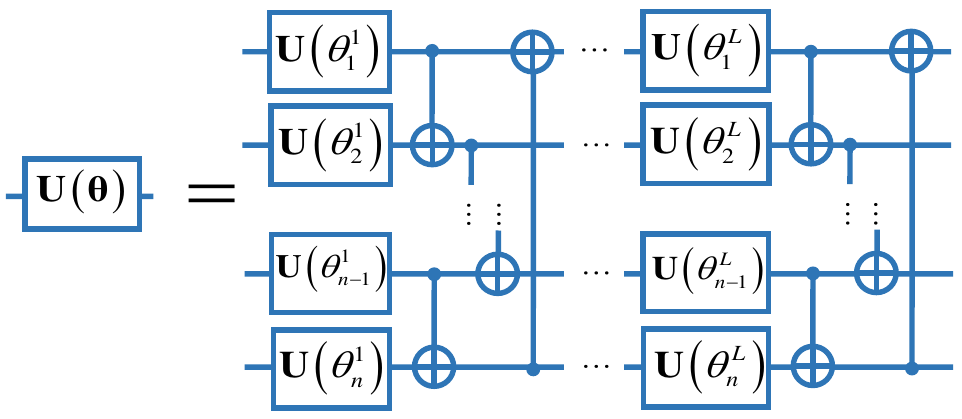}
\caption{Schematic diagram of parameterized quantum circuit $\mathcal{U}(\boldsymbol\theta)$.}
\end{figure}

\textbf{Theorem 1.} The gradients of loss functions $\mathcal{F}(\boldsymbol\theta)$ and $\mathcal{F}_j(\boldsymbol\theta)$ can be estimated on near-term quantum devices if and only if the operators
$\frac{\partial\mathcal{U}(\boldsymbol\theta)}{\partial\theta_i^t}$ and
$\frac{\partial\mathcal{U}^{\dag}(\boldsymbol\theta)}{\partial\theta_i^t}$
can be efficiently implemented on near-term quantum computers. Particularly, the derivatives of $\mathcal{U}(\boldsymbol\theta)$ and $\mathcal{U}^{\dag}(\boldsymbol\theta)$ with respect to a certain angle $\theta_i^t$ can be obtained using the following formulae,
$$\begin{aligned}
\frac{\partial\mathcal{U}(\boldsymbol\theta)}{\partial\theta_i^t}=\frac{1}{2}
\mathcal{U}(\boldsymbol\theta_{+}),~~~
\frac{\partial\mathcal{U}^{\dag}(\boldsymbol\theta)}{\partial\theta_i^t}
=\frac{1}{2}\mathcal{U}^{\dag}(\boldsymbol\theta_{+}),
\end{aligned}$$
where $\boldsymbol\theta=(\theta_1^1,\cdots,\theta_i^t,\cdots)^{\dag}$ and $\boldsymbol\theta_{+}=(\theta_1^1,\cdots,\pi+\theta_i^t,\cdots)^{\dag}$.

Here we discuss how to choose valid parameter $\gamma_i$ in loss function $\mathcal{F}_j(\boldsymbol\theta)$ . Let $|\psi(\boldsymbol\theta)\rangle:=\sum_{i=1}^{r}a_i|\psi_i\rangle$ be the trial state, where $|\psi_i\rangle$ is the $i$th generalized eigenvector. We can choose $\gamma_i=\lambda_{r}-\lambda_1>\lambda_j-\lambda_i$, where $\lambda_{r}$ and $\lambda_1$ can be estimated by optimizing the loss function $\mathcal{F}(\boldsymbol\theta)$. It can be easily verified that
\begin{equation}
\begin{aligned}
\mathcal{F}_j(\boldsymbol\theta)&=\frac{\sum_{i=1}^{r}|a_i|^2\lambda_i}{\sum_{i=1}^{r}|a_i|^2}+\sum_{i=1}^{j-1}\gamma_i|a_i|^2\\
&=\sum_{i=j}^{r}|a_i|^2\lambda_i+\sum_{i=1}^{j-1}(\gamma_i+\lambda_i)|a_i|^2\\
&\geq\lambda_j\sum_{i=1}^{r}|a_i|^2=\lambda_j,
\end{aligned}
\end{equation}
where the first equality is due to that $\sum_{i=1}^{r}|a_i|^2=1$ and $\langle\psi_i|\mathcal{B}|\psi_j\rangle=\delta_{ij}$, the first inequality follows as $\gamma_i+\lambda_i>\lambda_j$.
\subsection{Variational quantum generalized eigensolver}

The inputs of VQGE are a parameterized quantum circuit $\mathcal{U}(\boldsymbol\theta)$ and an initial state $|\psi_{\textrm{in}}\rangle$. The outputs of our algorithm are the generalized eigenvalues $\{\lambda_i\}_{i=1}^{r}$ and the optimal parameter $\{\boldsymbol\theta_i^{*}\}_{i=1}^{r}$ that prepares the eigenvectors $\{|\psi_i\rangle=\mathcal{U}(\boldsymbol\theta_i^{*})|\psi_{\textrm{in}}\rangle\}$. Here we summarize the iteration process of finding the generalized eigenvalues and eigenvectors.

(1) Compute the loss functions $\mathcal{F}(\boldsymbol\theta)$ and $\mathcal{F}_j(\boldsymbol\theta)$ on near-term computers.

(2) Apply the classical optimization procedure on loss functions $\mathcal{F}(\boldsymbol\theta)$ and $\mathcal{F}_j(\boldsymbol\theta)$ with the help of Theorem 1 and determine the optimal parameters $\boldsymbol\theta_1^{*},\boldsymbol\theta_{r}^{*},\boldsymbol\theta_j^{*}$.

(3) We obtain the $j^{\textrm{th}}$ generalized eigenvalues, $\lambda_j=\mathcal{F}_{j}(\boldsymbol\theta_j^{*})$ and eigenvectors $|\psi_j\rangle=|\psi(\boldsymbol\theta_j^{*})\rangle=\mathcal{U}(\boldsymbol\theta_j^{*})|\psi_{\textrm{in}}\rangle$. The lowest (largest) generalized eigenvalue $\lambda_1=\mathcal{F}(\boldsymbol\theta_1^{*})$ ($\lambda_r=\mathcal{F}(\boldsymbol\theta_r^{*})$) with eigenvectors $|\psi_1\rangle=|\psi(\boldsymbol\theta_1^{*})\rangle$ ($|\psi_r\rangle=|\psi(\boldsymbol\theta_r^{*})\rangle$), respectively.

Concerning the related errors and sample costs, we have the following theorem.

\textbf{Theorem 2.} Let $\epsilon_{\mathcal{A}}$, $ \epsilon_{\mathcal{B}}$ and $\epsilon_{\mathcal{O}}$ denote the corresponding precision of estimating $\langle\mathcal{A}\rangle$, $\langle\mathcal{B}\rangle$ and $\textrm{Tr}[\mathcal{B}\mathcal{U}(\boldsymbol\theta)\rho\mathcal{U}^{\dag}
(\boldsymbol\theta_i^{*})]$. For the GE problem (\ref{GEP}), there exists a quantum generalized eigensolver that outputs the former $j$ generalized eigenvalues with error $\eta_{1}^{-1}(\epsilon_{\mathcal{A}}+|\lambda_r|\epsilon_{\mathcal{B}})+\epsilon_{\mathcal{O}}$, where $\eta_1$ denotes the lowest eigenvalue of matrix $\mathcal{B}$ and $\lambda_{r}$ is the largest generalized eigenvalue. The total sample complexity is $O(j\Lambda^2\epsilon^{-2})$, $\Lambda=\sum_{k=0}^{K-1}\alpha_k+\sum_{l=0}^{L-1}\beta_l+\lambda_{r}$, $\epsilon^{2}=\epsilon_{\mathcal{A}}^2+\epsilon_{\mathcal{B}}^2+\epsilon_{\mathcal{O}}^2$.

The proof of Theorem 2 is provided in Appendix B.

{\it Remark:} In principle, VQGE can obtain the $r$ distinct generalized eigenvalues by optimizing $r$ loss functions. However, in general $r$ will be exponential in the number of qubits, so actually VQGE (including any classical algorithm) to find all of the generalized eigenvalues will take exponential time. In practice we are usually not interested in all of the eigenvalues, but only in some polynomial-sized subset of either the highest or lowest generalized eigenvalues. For example, the largest generalized eigenvector of covariance matrices $S_1$ and $S_2$ is referred to as the Fisher direction or Fisher axis \cite{fisher1936the}. One then can maximize the projection variance of means of classes and minimize the projection variance of class instances.

A recent work by Parker \emph{et al.} \cite{parker2020quantum} demonstrates that the GE problem can be solved efficiently as a standard Hermitian eigenvalue problem by introducing the immediate matrix $\mathcal{B}^{1/2}$ or the Cholesky decomposition of $\mathcal{B}$. Then one can apply the quantum phase estimation to solve the standard Hermitian eigenvalue problem, $\mathcal{B}^{-1/2}\mathcal{A}\mathcal{B}^{-1/2}|\Psi\rangle=\lambda|\Psi\rangle$, where $|\Psi\rangle=\mathcal{B}^{1/2}|\psi\rangle$ \cite{abrams1999quantum}. However, QPE requires fully coherence time and millions of quantum gates for practical application \cite{jones2012faster}, as compared to the $\mathcal{O}(2L\log N)$ quantum gates in our algorithm, where $L$ is a constant. Moreover, in \cite{parker2020quantum} $\mathcal{A}$ is assumed to be local and sparse and $\mathcal{B}$ is supposed to be sparse. However, VQGE does not need any assumption on the structure of Hermitian matrices $\mathcal{A}$ and $\mathcal{B}$.

In Reference \cite{liang2019variational}, the original matrices $\mathcal{A}$ and $\mathcal{B}$ are transformed into a quadratic form, $(\mathcal{A}-\tau\mathcal{B})^2$ and $\mathcal{B}^2$ respectively by introducing an extra variable $\tau$. One needs to tune the parameter $\tau$ to get an approximate generalized eigenvalue, which results in much more resources than our algorithm. Furthermore, the quadratic Hamiltonian increases the resulting errors as extra matrix multiplications costs more computation resources.

If the entire quantum algorithm is performed on a classical computer, our algorithm attains an exponential advantage in the computation of loss function and the storage of quantum states. Namely, for a $\log N$-qubit quantum state $|\psi\rangle$, current classical algorithm requires $N$ complex numbers. Even if the quantum state is prepared perfectly, the computation of the expectation value $\langle\mathcal{A}\rangle$ or $\langle\mathcal{B}\rangle$ requires $\mathcal{O}(dN)$ floating point operations, where $d$ is the number of Pauli terms of matrices $\mathcal{A}$ and $\mathcal{B}$. Thus we require exponential resources in both storage and computation when performed on a classical computer.
\section{Iteration method: a full quantum generalized eigensolver}
This section presents an alternate quantum algorithm to obtain the minimum generalized eigenvalue and corresponding eigenvectors on a fault-tolerant quantum computer. We call this iteration algorithm a full quantum generalized eigensolver (FQGE), a iterative quantum algorithm, which does not require quantum phase estimation. It is notice that the classical optimization procedure of VQGE is replaced by a quantum gradient descent algorithm \cite{wei2020a,li2021optimizing}.
\subsection{Theoretical basis of FQGE}
For the GE problem (\ref{GEP}) considering the loss function
\begin{equation}
\begin{aligned}
\mathcal{F}(|\psi\rangle)=\frac{\langle\psi|\mathcal{A}|\psi\rangle}{\langle\psi|\mathcal{B}|\psi\rangle},
\end{aligned}
\end{equation}
we optimize $\mathcal{F}(|\psi\rangle)$ by the recently proposed quantum gradient descent algorithm \cite{wei2020a,li2021optimizing}. The basic idea is to construct a sequence $\{|\psi_s\rangle\in\mathbb{R}^{N\times N}\}_{s=1,2,\cdots}$ such that $\mathcal{F}(|\psi_{s+1}\rangle)<\mathcal{F}(|\psi_{s}\rangle)$ for all $s$. If the iterative is terminated under a given termination condition, the sequence $\mathcal{F}(|\psi_{s}\rangle)$ converges to the minimal eigenvalue $\lambda_1$ and the state sequence $\{|\psi_s\rangle\}$ to the corresponding eigenvector.

For any given $|\psi_{s}\rangle$, the next state is updated using the following transformation
\begin{equation}\label{iterative}
\begin{aligned}
|\psi_{s+1}\rangle&=|\psi_{s}\rangle+\delta|\widetilde{\psi}_{s}\rangle\\
&=|\psi_{s}\rangle-\delta\nabla\mathcal{F}(|\psi_{s}\rangle)|\psi_{s}\rangle,
\end{aligned}
\end{equation}
where $\delta$ is the learning rate and we choose the search direction of $|\widetilde{\psi}_{s}\rangle$ to be the negative gradient of the cost function $\mathcal{F}(|\psi_{s}\rangle)$. The parameter $\delta$ is determined such that the function value $\mathcal{F}(|\psi_{s+1}\rangle)$ of the new state $|\psi_{s+1}\rangle$ becomes minimal,
\begin{equation}\label{learningrate}
\begin{aligned}
\mathcal{F}(|\psi_{s+1}\rangle)=\min_{\delta\in\mathbb{C}}\mathcal{F}(|\psi_{s}\rangle+\delta|\widetilde{\psi}_{s}\rangle).
\end{aligned}
\end{equation}
The minimal in Eq. (\ref{learningrate}) is the smaller generalized eigenvalue $u_1$ of the generalized $2\times 2$ eigenvalue problem
$$\begin{bmatrix}
\langle\psi_{s}|\mathcal{A}|\psi_{s}\rangle & \langle\psi_{s}|\mathcal{A}|\widetilde{\psi}_{s}\rangle\\
\langle\widetilde{\psi}_{s}|\mathcal{A}|\psi_{s}\rangle & \langle\widetilde{\psi}_{s}|\mathcal{A}|\widetilde{\psi}_{s}\rangle
\end{bmatrix}
|u\rangle=u
\begin{bmatrix}
\langle\psi_{s}|\mathcal{B}|\psi_{s}\rangle & \langle\psi_{s}|\mathcal{B}|\widetilde{\psi}_{s}\rangle\\
\langle\widetilde{\psi}_{s}|\mathcal{B}|\psi_{s}\rangle & \langle\widetilde{\psi}_{s}|\mathcal{B}|\widetilde{\psi}_{s}\rangle
\end{bmatrix}
|u\rangle.
$$
The corresponding eigenvector is normalized such that its component equals $|u_1\rangle=[1,\delta]^{T}$.

The gradient operator of the loss function is
\begin{equation}
\begin{aligned}
\nabla\mathcal{F}(|\psi\rangle)&=\frac{2[\mathcal{A}-\mathcal{F}(|\psi\rangle)\mathcal{B}]}{\langle\psi|\mathcal{B}|\psi\rangle}.
\end{aligned}
\end{equation}
Then the iterative Eq.(\ref{iterative}) is interpreted as an evolution with non-unitary operator $G_{s}$,
\begin{equation}
\begin{aligned}
&|\psi_{s+1}\rangle=G_{s}|\psi_{s}\rangle,\\
&G_{s}=I-\frac{2\delta[\mathcal{A}-\mathcal{F}(|\psi_{s}\rangle)\mathcal{B}]}{\langle\psi_{s}|\mathcal{B}|\psi_{s}\rangle}.
\end{aligned}
\end{equation}
Based on the decomposition of matrices $\mathcal{A}$ and $\mathcal{B}$, The non-unitary operator $G_{s}$ can also be decomposed into a linear combination of unitary operators (LCU) \cite{long2006general}, such that
\begin{equation}
\begin{aligned}
G_{s}=\sum_{i=0}^{d-1}g_{s,i}G_{s,i}.
\end{aligned}
\end{equation}
Given the decomposition (\ref{GEP}), the Pauli product terms $G_{s,i}\in\{I,A_k,B_l\}$ and coefficients $$g_{s,i}\in\Big\{1,-\frac{2\delta\alpha_k}{\langle\psi_{s}|\mathcal{B}|\psi_{s}\rangle},
-\frac{2\delta\mathcal{F}(|\psi_{s}\rangle)\beta_l}{\langle\psi_{s}|\mathcal{B}|\psi_{s}\rangle}\Big\}$$
for all $k$, $l$. The number of Pauli product terms $G_{s,i}$ is $d=K+L+1=O(poly\log N)$. The Hadamard test enables us to compute the coefficients $g_{s,i}$. The approach proposed in Ref. \cite{berry2015simulating} is used to implement the non-unitary operator $G_{s}$. It is clear that $G_{s}$ is $|\psi_{s}\rangle$ dependent.

The termination condition is to check if
\begin{equation}
\begin{aligned}
\frac{\|\mathcal{A}|\psi_{s+1}\rangle-\mathcal{F}(|\psi_{s+1}\rangle)\mathcal{B}|\psi_{s+1}\rangle\|_2}
{\|\mathcal{A}|\psi_{s+1}\rangle\|_2+|\mathcal{F}(|\psi_{s+1}\rangle)|\|\mathcal{B}|\psi_{s+1}\rangle\|_2}\leq\varepsilon
\end{aligned}
\end{equation}
where $\varepsilon$ is a given tolerance \cite{arbenz2016solving}. If it is satisfied, $|\psi_{s+1}\rangle$ will be thought to be a eigenvector associated with inferred generalized eigenvalue $\lambda_1=\mathcal{F}(|\psi_{s+1}\rangle)$.

As pointed out in \cite{arbenz2016solving}, our FQGE converges slowly. This happens if the generalized spectrum is very much spread out, i.e., if the condition number of $\mathcal{A}$ relative to $\mathcal{B}$ is big. Note that the perturbation bounds for generalized eigenvalue problems is the same as the work \cite{golub2002an}.
\subsection{The construction of quantum circuit}
The quantum circuit of one iteration process is shown in Fig. 2. Firstly, the superposition state $C^{-1}\sum_{i=0}^{d-1}g_{s,i}|i\rangle$ should be efficiently prepared, where $C=\sqrt{\sum_{i=0}^{d-1}g_{s,i}^2}$ is a normalization constant and $|i\rangle$ is the computational basis. If state preparation is too expensive, it can diminish the quantum advantage. Some methods is available for the state preparation, such as approaches based on Grover search \cite{long2001efficient,vazquez2001efficient} and quantum random access memory (qRAM) \cite{soklakov2006quantum,giovannetti2008quantum}. In general applying unitary
\begin{equation}
\begin{aligned}
W=\frac{1}{C}
\begin{bmatrix}
g_{s,0} & w_{0,1} & \cdots & w_{0,d-1}\\
g_{s,1} & w_{1,1} & \cdots & w_{1,d-1}\\
\cdots & \cdots & \cdots & \cdots\\
g_{s,d-1} & w_{d-1,1} & \cdots & w_{d-1,d-1}
\end{bmatrix}
\end{aligned}
\end{equation}
on $\bar{d}=\log d\in\mathcal{O}(\log poly\log N)$ ancillary state $|0\cdots0\rangle$. The elements $\{w_{0,1},\cdots,w_{d-1,d-1}\}$ are arbitrary as long as $W$ is unitary. Moreover, we prepare an initial state $|\psi_{0}\rangle$ in a $\log N$-qubit system. In our experiment, $|\psi_{0}\rangle$ usually is chosen as some easily prepared state, such as a tensor product state $|0\cdots0\rangle$. Thus, the state of the whole system is
\begin{equation}
\begin{aligned}
|\Psi_0\rangle=\frac{1}{C}\sum_{i=0}^{d-1}g_{s,i}|i\rangle|\psi_{s}\rangle.
\end{aligned}
\end{equation}

\begin{figure}[ht]
\includegraphics[scale=1]{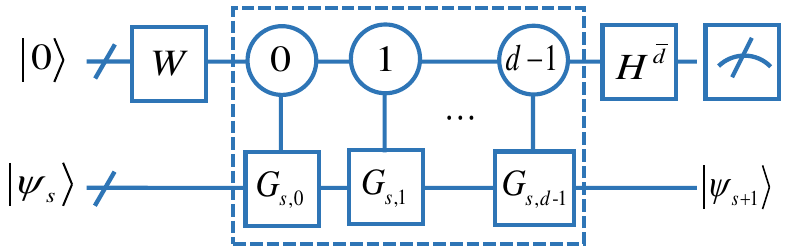}
\caption{Quantum circuit of FQGE.}
\end{figure}

Next, $\bar{d}$-qubit-controlled quantum gates
\begin{equation}
\begin{aligned}
C_{\bar{d}}(G_s)=\sum_{i=0}^{d-1}|i\rangle\langle i|\otimes G_{s,i}
\end{aligned}
\end{equation}
are implemented on state $|\Psi_0\rangle$ and we obtain state
\begin{equation}
\begin{aligned}
|\Psi_1\rangle&=C_{\bar{d}}(G_{s,i})|\Psi_0\rangle\\
&=\frac{1}{C}\Bigg(\sum_{i=0}^{d-1}g_{s,i}|i\rangle G_{s,i}|\psi_{s}\rangle\Bigg).
\end{aligned}
\end{equation}
Based on the operator $G_{s,i}$ is a tensor product of Pauli operators, controlled operation $C_{\bar{d}}(G_{s})$ can be decomposed into a sequence of controlled operation $C_{\bar{d}}(G_{s,i})=|i\rangle\langle i|\otimes G_{s,i}$ with $i$ running from 0 to $d-1$. Each individual operation $C_{\bar{d}}(G_{s,i})$ can be further transformed into some multi-qubit controlled Pauli gate which is simulated by a network consisting of elementary ($1$- and $2$-qubit) gates using the Lemma 7.5 introduced in \cite{barenco1995elementary}. Let $M_{s,i}$ be such that $M_{s,i}^2=G_{s,i}$. The unitary $C_{\bar{d}}(G_{s,i})$ can be replaced by two $\bar{d}$-qubit Toffoli gates, $C_{1}(M_{s,i})$, $C_{1}(M_{s,i}^{\dag})$, and one controlled operation $C_{\bar{d}-1}(M_{s,i})$. The cost of simulating the 2 $\bar{d}$-qubit Toffoli gates, $C_{1}(M_{s,i})$ and $C_{1}(M_{s,i}^{\dag})$ are $\mathcal{O}(\bar{d})$, $\mathcal{O}(\log N)$ and $\mathcal{O}(\log N)$, respectively. Let $T_{\bar{d},\log N}$ be the total gate complexity of simulating $C_{\bar{d}}(G_{s,i})$. We obtain the recursion equation,
\begin{equation}
\begin{aligned}
T_{\bar{d},\log N}&=T_{\bar{d}-1,\log N}+O(\log N)+O(\bar{d})\\
&=\mathcal{O}(\log d\log N+\log d).
\end{aligned}
\end{equation}
Thus, $d$ unitary gates $\{C_{\bar{d}}(G_{s,i})\}$ can be decomposed into $\mathcal{O}(d\log d\log N+d\log d)$ elementary quantum gates. It is clear that the complexity of quantum gate is polylogarithmical to the size of matrix $G_{s}$. However, the classical counterparts have complexity $\mathcal{O}(N^2)$. Therefore, in comparison, FQGE achieves an exponential speedup compared in the problem size $N$ under assumption decomposition of $\mathcal{A}$ and $\mathcal{B}$.

After performing $\bar{d}$ Hadamard gates on ancillary register, we transform state $|\Psi_1\rangle$ into state
\begin{equation}
\begin{aligned}
|\Psi_2\rangle&=\frac{1}{C}\Bigg(\sum_{i=0}^{d-1}g_{s,i}H^{\otimes\bar{d}}|i\rangle G_{s,i}|\psi_{s}\rangle\Bigg)\\
&=\frac{1}{C\sqrt{d}}|0\cdots0\rangle\Bigg(\sum_{i=0}^{d-1}g_{s,i}G_{s,i}|\psi_{s}\rangle\Bigg).
\end{aligned}
\end{equation}

Measuring the ancillary register and obtaining result $|0\cdots0\rangle$, the collapsed state can be viewed as the updated state $|\psi_{s+1}\rangle=\frac{1}{C\sqrt{d}}\sum_{i=0}^{d-1}g_{s,i}G_{s,i}|\psi_{s}\rangle.$ The success probability of obtaining $|0\cdots0\rangle$ is
\begin{equation}
\begin{aligned}
P_{\textrm{suc}}=\frac{\|G_s|\psi_{s}\rangle\|^2}{C^2d}.
\end{aligned}
\end{equation}
The measurement complexity is $C^2d/\|G_s|\psi_{s}\rangle\|^2$. The fact that FQGE requires a measurement at the end of each iterative step. the probability for success of each step is at least $\mathcal{O}(1/poly\log N)$, the FQGE needs to converge to the lowest generalized eigenvalue in only polylogarithmically-many steps. Then there is hope for the FQGE to be efficient. The total number of qubits is $\mathcal{O}(\log poly\log N)$ which is less than the qubit cost of standard VQE.

\section{Numerical results}
We first apply our algorithm, VQGE, to solve a GE problem with two 2-qubit Hermitian matrices. The simulation and optimization loops are carried out via Paddle Quantum \cite{paddlequantum} on the PaddlePaddle Deep Learning Platform \cite{paddle2}. We consider the following GE problem: $\mathcal{A}|\psi\rangle=\lambda\mathcal{B}|\psi\rangle$, with
$$
\begin{aligned}
&\mathcal{A}=\mathds{1}_4+0.4Z\otimes \mathds{1}_2+0.4\mathds{1}_2\otimes Z+0.2X\otimes X,\\
&\mathcal{B}=\mathds{1}_4+0.3Z\otimes \mathds{1}_2+0.4\mathds{1}_2\otimes Z+0.2Z\otimes Z,
\end{aligned}
$$
where $X$ and $Z$ are the standard Pauli operators, $\mathds{1}_d$ denotes the $d\times d$ identity matrix. We use a quantum circuit with $L=2$ layers with initial state $|\psi_{\textrm{in}}\rangle=|0\rangle\otimes|0\rangle$. Our algorithm finds all $4$ generalized eigenvalues and corresponding eigenvectors. Notice that hardware noise is not considered in this heuristics.

For this case, the generalized eigenvalues of the matrix pair $(\mathcal{A},\,\mathcal{B})$ is non-degenerate. Fig. 3 plots the values of loss functions versus the number of iterations (with the total number of iterations fixed). As shown in Fig. 3, for the smallest eigenvalue, the loss function reaches a minimal value corresponding to an inferred generalized eigenvalue $\lambda_1$. The top line shows that the maximal value of function $\mathcal{F}(\boldsymbol\theta)$ approximately represents the fourth eigenvalue $\lambda_4$. It is clear that the minimal and the maximal values of the function $\mathcal{F}(\boldsymbol\theta)$ are attained by increasing iterations. However, for the second eigenvalues, the function escapes from the global minimal value with the increase of the iteration. As shown in Fig. 3, the global minimal value indicated by the red triangle can be reached at $17$th step. Therefore, by recording the iteration process, we can search the global minimal value in the experimental result set and obtain the associated parameter $\boldsymbol\theta_{\textrm{opt}}$.
\begin{figure}[ht]
\includegraphics[scale=0.5]{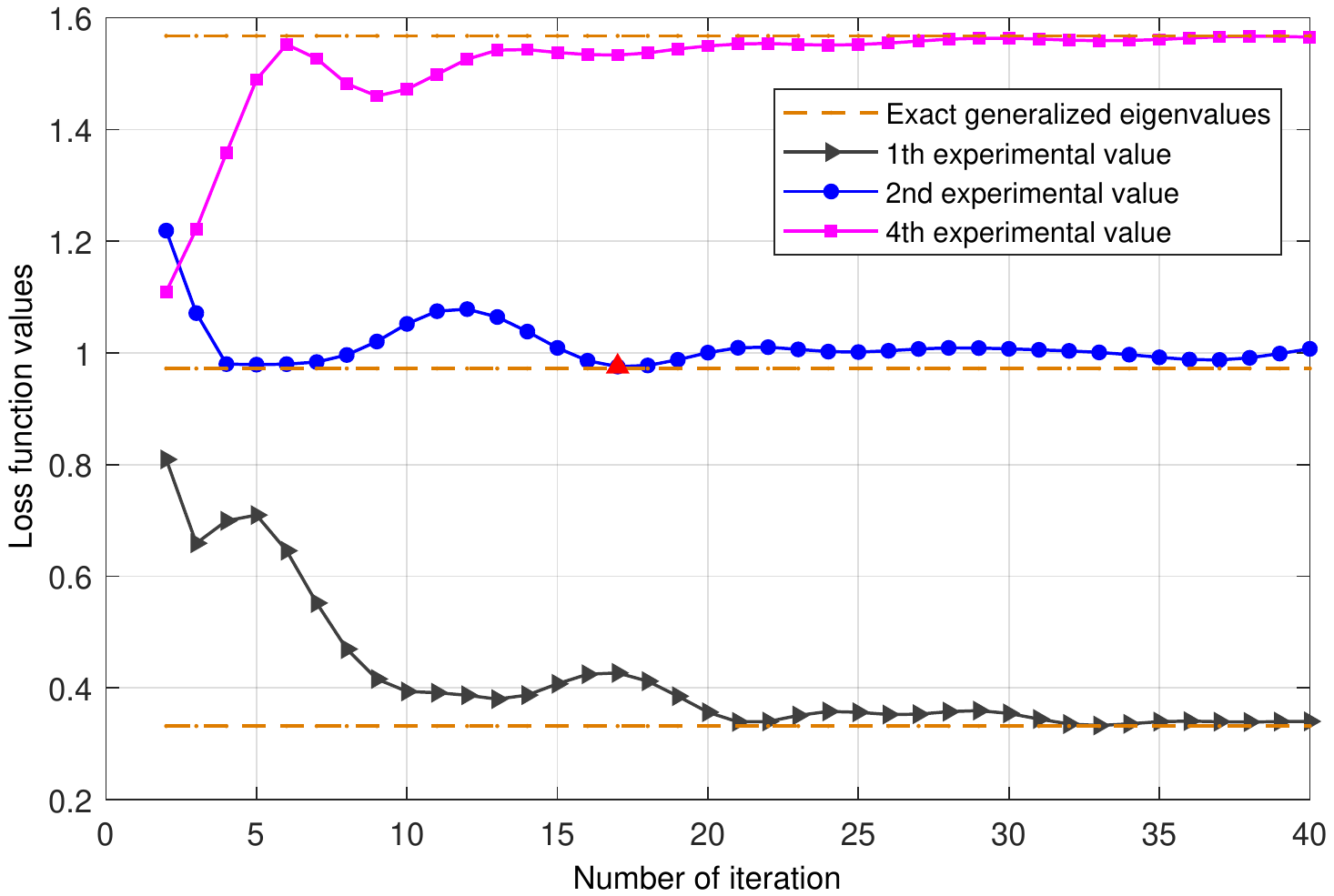}
\caption{Iteration process for finding generalized eigenvalues of matrix pair ($\mathcal{A},\mathcal{B}$). Here, we choose the circuit depth as $L=2$. The minimum of two loss functions implies an approximate generalized eigenvalue. And the optimal parameter $\boldsymbol\theta_{j}^{*}$ enables us to prepare the eigenvector $|\psi(\boldsymbol\theta_{j}^{*})\rangle$.}
\end{figure}

Here, we focus on the calculation of the minimum generalized eigenvalue of $(\mathcal{A},\mathcal{B})$ using FQGE. This numerical simulation is made with a classical computer that simulates the quantum iterative processes and introduces Gaussian noise. In our two-qubits system, the initial state $|\psi_{0}\rangle=|00\rangle$ and the learning rate is chosen as $\delta=0.1$. As shown in Fig. 4, the experimental values converge to the exact eigenvalue with state fidelity $99.99\%$. We also investigate the influence of noise in Fig. 4. The Gaussian noise term
\begin{equation}
\begin{aligned}
|\psi_{\textrm{noise}}\rangle=\tau\sum_{i=0}^{3}\frac{1}{2}|i\rangle,
\end{aligned}
\end{equation}
is added to the iterative state $|\psi_{s}\rangle:=|\psi_{s}\rangle+|\psi_{\textrm{noise}}\rangle$, where the variable $\tau\sim\mathcal{N}(\mu,\sigma^2)$ follows a Gaussian distribution. In our experiment, we set $\mu=0$ and $\sigma=0.01$. The experiment values still converge to the exact eigenvalue. Thus our FQGE is robust to Gaussian noise.
\begin{figure}[ht]
\includegraphics[scale=0.5]{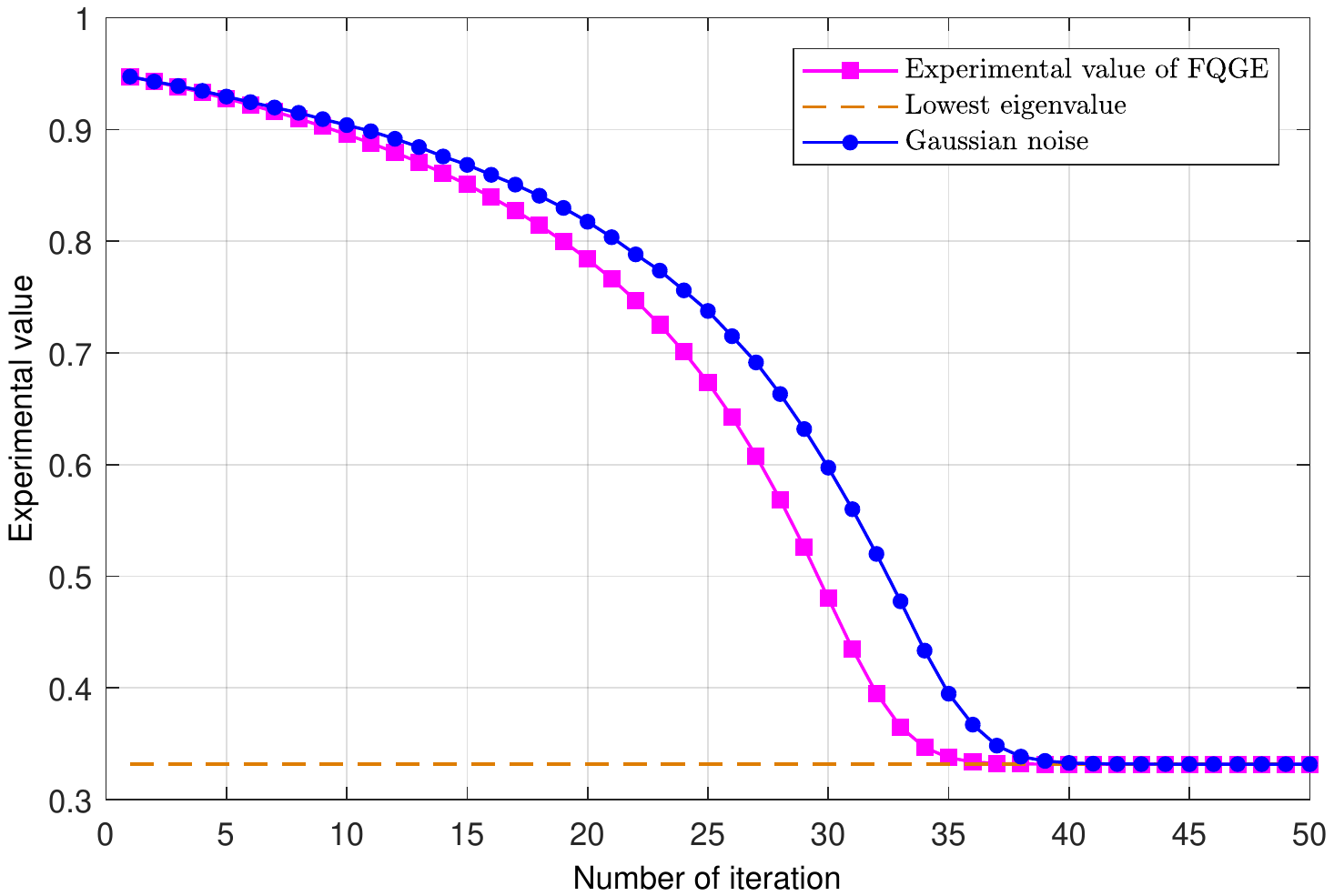}
\caption{The convergence to the lowest eigenvalue by the FQGE. The red line indicates the experimental value with Gaussian noise. The blue line is the exact eigenvalue. The pink line shows the experimental value without any type of noise.}
\end{figure}
\section{Conclusion}
In conclusion, we have proposed an efficient variational quantum algorithm, VQGE, for solving the generalized eigenvalues and eigenvectors of a given matrix pair $(\mathcal{A},\,\mathcal{B})$. It has been shown that the proposed FQGE can calculate the lowest generalized eigenvalue using quantum gradient descent. The number of elementary gates is polylogarithmical to the number of qubits. Both VQGE and FQGE theoretically achieve exponential speedup over its classical part under the efficient Pauli decomposition of matrix pair ($\mathcal{A},\mathcal{B}$). As our algorithm does not use the Hamiltonian simulation, amplitude amplification and phase estimation, it may have more efficient applications in such as dimensionality reduction \cite{bishop2006pattern}, molecular orbital computation \cite{huggins2020a,sakurai2010a} and electronic structure calculations \cite{teng2011efficient}.

Several basic mathematical problems left to be treated with. The first one is what type of variational circuit $\mathcal{U}(\boldsymbol\theta)$ should be designed for various practical problems. Different variational ansatz may affect the convergence property of the corresponding loss functions \cite{woitzik2020entanglement}. Moreover, although the problem of barren plateaus \cite{cerezo2020variational} has been solved  by defining an adaptive loss function, the phenomenon of barren plateaus introduced in \cite{mcclean2018barren} need to be further studied, as escaping the barren plateaus can guarantee that VQAs provide a speedup over classical algorithm. Finally, focusing on the FQGE, two caveats may make FQGE unsuitable for NISQ devices. The first one is the preparation of entanglement state $C^{-1}\sum_{i=0}^{d-1}g_{s,i}|i\rangle$. In the main text, we assume that applying unitary $W$ on initial $|0\cdots0\rangle$ can prepare the state. However, construct unitary $W$ is a hard question. Second potential question is the post measurement. The success probability decaying exponential with the number of iterations does not ensure us to obtain ground state with high probability. Although the amplitude amplification can amplify the amplitude up to a deterministic order, additional resources are need such as repetition of the overall procedure. The investigation on these related questions may shed new light on the quantum advantages in machine learning and artificial intelligence.

\bigskip

Acknowledgements: This work is supported by the NSF of China under Grant No. 12075159 and 11775306, Shenzhen Institute for Quantum Science and Engineering, Southern University of Science and Technology (Grant No.SIQSE202001), the Key Project of Beijing Municipal Commission of Education (Grant No. KZ201810028042), Beijing Natural Science Foundation (Z190005), the Academician Innovation Platform of Hainan Province, and Academy for Multidisciplinary Studies, Capital Normal University.

\appendix

\section{Proof of Theorem 1}
A unitary circuit $\mathcal{U}(\boldsymbol\theta)$ parameterized by $\boldsymbol\theta$ has the form,
\begin{equation}
\mathcal{U}(\boldsymbol\theta)=\Pi_{t=L}^1\mathcal{U}_{ent}
\Bigg(\otimes_{i=1}^{n}R_y(\theta_i^t)\Bigg),
\end{equation}
where $R_y(\theta_i^t)=e^{\frac{\textrm{i}}{2}\theta_i^tY}$ and $\boldsymbol\theta=(\theta_1^1,\cdots,\theta_1^L,\cdots,\theta_n^1,\cdots,\theta_n^L)^{\dag}$. Thus the derivative of $\mathcal{U}(\boldsymbol\theta)$ with respect to a certain angle $\theta_i^t$ is given by
\begin{equation}\label{partialu}
\begin{aligned}
\frac{\partial\mathcal{U}(\boldsymbol\theta)}{\partial\theta_i^t}
&=\Pi_{t=L}^1\mathcal{U}_{ent}\otimes_{i=1}^{n}\frac{\partial R_y(\theta_i^t)}{\partial\theta_i^t}\\
&=\Pi_{t=L}^1\mathcal{U}_{ent}\otimes_{i=1}^{n}\frac{-\textrm{i}Y}{2}R_y(\theta_i^t)\\
&=\frac{1}{2}\mathcal{U}(\boldsymbol\theta_{+}),
\end{aligned}
\end{equation}
where the last equality is true due to the fact that
\begin{equation}\label{pi}
\begin{aligned}
R_y(\pm\pi)&=e^{\mp\frac{\textrm{i}}{2}\pi Y}=\cos\frac{\pi}{2}I\mp\textrm{i}\sin\frac{\pi}{2}Y=\mp\textrm{i}Y
\end{aligned}
\end{equation}
and $\boldsymbol\theta_{+}=(\theta_1^1,\cdots,\pi+\theta_i^t,\cdots)^{\dag}$. Similarly, we have
\begin{equation}\label{partialu2}
\begin{aligned}
\frac{\partial\mathcal{U}^{\dag}(\boldsymbol\theta)}{\partial\theta_i^t}
=\Bigg(\frac{\partial\mathcal{U}(\boldsymbol\theta)}{\partial\theta_i^t}\Bigg)^{\dag}
=\frac{1}{2}\mathcal{U}^{\dag}(\boldsymbol\theta_{+}).
\end{aligned}
\end{equation}

Concerning the first loss function
\begin{align}
\mathcal{F}(\boldsymbol\theta)
=\frac{\langle\psi(\boldsymbol\theta)|\mathcal{A}|\psi(\boldsymbol\theta)\rangle}{\langle\psi(\boldsymbol\theta)|\mathcal{B}|\psi(\boldsymbol\theta)\rangle}
=\frac{\langle\mathcal{A}\rangle}{\langle\mathcal{B}\rangle},
\end{align}
we obtain
\begin{align}
\frac{\partial\mathcal{F}(\boldsymbol\theta)}{\partial\theta_i^t}
&=\frac{\partial}{\partial\theta_i^t}\frac{\langle\mathcal{A}\rangle}{\langle\mathcal{B}\rangle}\nonumber\\
&=\frac{1}{\langle\mathcal{B}\rangle^2}\Bigg(\frac{\partial\langle\mathcal{A}\rangle}{\partial\theta_i^t}\langle\mathcal{B}\rangle-\langle\mathcal{A}\rangle
\frac{\partial\langle\mathcal{B}\rangle}{\partial\theta_i^t}\Bigg).
\end{align}
It is clear that the inner product items $\langle\mathcal{A}\rangle$ and $\langle\mathcal{B}\rangle$ can be calculated on NISQ quantum devices by applying the Hadamard test or the Theorem 1 introduced in the main text. Furthermore, two extra items $\frac{\partial\langle\mathcal{A}\rangle}{\partial\theta_i^t}$ and $\frac{\partial\langle\mathcal{B}\rangle}{\partial\theta_i^t}$ can be transformed into computational-friendly forms. In particular,
\begin{equation}\label{partiala}
\begin{aligned}
\frac{\partial\langle\mathcal{A}\rangle}{\partial\theta_i^t}
=\textrm{Tr}[\mathcal{A}\mathcal{U}(\boldsymbol\theta)\rho\frac{\partial\mathcal{U}^{\dag}(\boldsymbol\theta)}{\partial\theta_i^t}]+
\textrm{Tr}[\mathcal{A}\frac{\partial\mathcal{U}(\boldsymbol\theta)}{\partial\theta_i^t}\rho\mathcal{U}^{\dag}(\boldsymbol\theta)].
\end{aligned}
\end{equation}
Substituting (\ref{partialu}) and (\ref{partialu2}) into (\ref{partiala}), we get
\begin{equation}\label{partiala1}
\begin{aligned}
\frac{\partial\langle\mathcal{A}\rangle}{\partial\theta_i^t}
=\frac{1}{2}\Bigg(\textrm{Tr}[\mathcal{A}\mathcal{U}(\boldsymbol\theta)
\rho\mathcal{U}^{\dag}(\boldsymbol\theta_{+})]+
\textrm{Tr}[\mathcal{A}\mathcal{U}(\boldsymbol\theta_{+})
\rho\mathcal{U}^{\dag}(\boldsymbol\theta)]\Bigg).
\end{aligned}
\end{equation}
Similarly, we have
\begin{align}\label{partialb}
\frac{\partial\langle\mathcal{B}\rangle}{\partial\theta_i^t}
=\frac{1}{2}\Bigg(\textrm{Tr}[\mathcal{B}\mathcal{U}(\boldsymbol\theta)\rho\mathcal{U}^{\dag}(\boldsymbol\theta_{+})]+
\textrm{Tr}[\mathcal{B}\mathcal{U}(\boldsymbol\theta_{+})\rho\mathcal{U}^{\dag}(\boldsymbol\theta)]\Bigg).
\end{align}

To compute the derivative of the loss function $\mathcal{F}_j(\boldsymbol\theta)$, we study the derivative of the extra item $\sum_{i=1}^{j-1}\gamma_i\textrm{Tr}[\mathcal{B}\mathcal{U}(\boldsymbol\theta)
\rho\mathcal{U}^{\dag}(\boldsymbol\theta_i^{*})]^2$,
\begin{align}
&\frac{\partial}{\partial\theta_i^t}
\sum_{i=1}^{j-1}\gamma_i\textrm{Tr}[\mathcal{B}\mathcal{U}(\boldsymbol\theta)\rho\mathcal{U}^{\dag}(\boldsymbol\theta_i^{*})]^2\nonumber\\
&=2\sum_{i=1}^{j-1}\gamma_i\textrm{Tr}[\mathcal{B}\mathcal{U}(\boldsymbol\theta)\rho\mathcal{U}^{\dag}(\boldsymbol\theta_i^{*})]
\textrm{Tr}[\mathcal{B}\frac{\partial\mathcal{U}(\boldsymbol\theta)}{\partial\theta_i^t}\rho\mathcal{U}(\boldsymbol\theta_i^{*})]\nonumber\\
&=\sum_{i=1}^{j-1}\gamma_i\textrm{Tr}[\mathcal{B}\mathcal{U}(\boldsymbol\theta)\rho\mathcal{U}^{\dag}(\boldsymbol\theta_i^{*})]
\textrm{Tr}[\mathcal{B}\mathcal{U}(\boldsymbol\theta_{+})\rho\mathcal{U}(\boldsymbol\theta_i^{*})].\nonumber
\end{align}
Therefore, the gradient of loss function $\mathcal{F}_j(\boldsymbol\theta)$ is given by
\begin{equation}
\begin{aligned}
\frac{\partial\mathcal{F}_j(\boldsymbol\theta)}{\partial\theta_i^t}
&=\frac{\partial\mathcal{F}(\boldsymbol\theta)}{\partial\theta_i^t}+\frac{\partial}{\partial\theta_i^t}
\sum_{i=1}^{j-1}\gamma_i\textrm{Tr}[\mathcal{B}\mathcal{U}(\boldsymbol\theta)\rho\mathcal{U}^{\dag}(\boldsymbol\theta_i^{*})]^2.
\end{aligned}
\end{equation}

An alternative method to estimate the gradient is using the Finite-difference approximation,
\begin{align}
&\frac{\partial\mathcal{F}(\boldsymbol\theta)}{\partial\theta_i^t}
=\frac{\mathcal{F}(\boldsymbol\theta+\Delta\theta)-\mathcal{F}(\boldsymbol\theta-\Delta\theta)}{2\Delta\theta},
\end{align}
where $\Delta\theta$ is a perturbation on the $\boldsymbol\theta$. Usually, we fix $\Delta\theta$ at a sufficiently small value.

In a conclusion, the calculation of the gradient of the functions $\mathcal{F}(\boldsymbol\theta)$ and $\mathcal{F}_j(\boldsymbol\theta)$ is carried out on a NISQ quantum devices.

\section{Proof of Theorem 2}
The proof of Theorem 2 contains two parts: the error analysis and the sample complexity. First of all, the depth of variational circuit used in our algorithm is determined by the number of quantum gates applied on each qubit. Our variational circuit (Fig. 1 in the main text) has $2L$ quantum gates on each qubit if we repeat the block $L$ times. Thus the depth scales as $2L=\mathcal{O}(1)$ that is independent of the size of the GE problem.

\subsection{Error analysis}
The main error is dominated by the quantum measurements. Assume that the quantities
$\langle\mathcal{A}\rangle=\sum_{k=0}^{K-1}\alpha_k\langle\psi(\boldsymbol\theta)
|\mathcal{A}_k|\psi(\boldsymbol\theta)\rangle,$
$\langle\mathcal{B}\rangle=\sum_{l=0}^{L-1}\beta_l\langle\psi(\boldsymbol\theta)
|\mathcal{B}_l|\psi(\boldsymbol\theta)\rangle$ and $\sum_{i=1}^{j-1}\gamma_i|\langle\psi(\boldsymbol\theta)|\mathcal{B}
|\psi(\boldsymbol\theta_{i}^{*})\rangle|^2$
have errors $\epsilon_{\mathcal{A}}$, $\epsilon_{\mathcal{B}}$ and $\epsilon_{\mathcal{O}}$, respectively. Let $\tilde{\mathcal{F}}(\boldsymbol\theta)$ and $\tilde{\mathcal{F}}_j(\boldsymbol\theta)$ denote
the estimated values of the functions $\mathcal{F}(\boldsymbol\theta)$ and $\mathcal{F}_j(\boldsymbol\theta)$, respectively.

The error of the defined function $\mathcal{F}(\boldsymbol\theta)$ is given by
$$
\begin{aligned}
\mathcal{E}&=
\left|\tilde{\mathcal{F}}(\boldsymbol\theta)-\mathcal{F}(\boldsymbol\theta)\right|=\left|\frac{\langle\mathcal{A}\rangle\pm\epsilon_{\mathcal{A}}}
{\langle\mathcal{B}\rangle\pm\epsilon_{\mathcal{B}}}
-\frac{\langle\mathcal{A}\rangle}{\langle\mathcal{B}\rangle}\right|\\
&=\left|
\frac{\langle\mathcal{B}\rangle\epsilon_{\mathcal{A}}\mp\langle\mathcal{A}\rangle\epsilon_{\mathcal{B}}}
{\langle\mathcal{B}\rangle(\langle\mathcal{B}\rangle\pm\epsilon_{\mathcal{B}})}\right|
\end{aligned}$$
Using the triangle inequality, we get
$$
\begin{aligned}
\mathcal{E}&\leq\left|
\frac{\langle\mathcal{B}\rangle\epsilon_{\mathcal{A}}\mp\langle\mathcal{A}\rangle\epsilon_{\mathcal{B}}}{\langle\mathcal{B}\rangle^2}\right|
\leq\left|\frac{\epsilon_{\mathcal{A}}}{\langle\mathcal{B}\rangle}\right|+
\left|\frac{\langle\mathcal{A}\rangle}{\langle\mathcal{B}\rangle}\frac{\epsilon_{\mathcal{B}}}{\langle\mathcal{B}\rangle}\right|\\
&\leq\eta_{1}^{-1}(\epsilon_{\mathcal{A}}+|\lambda_r|\epsilon_{\mathcal{B}}),
\end{aligned}$$
where $\eta_{1}$ is the smallest eigenvalue of $\mathcal{B}$. Similarly, the error of objective function $\mathcal{F}_j(\boldsymbol\theta)$ is
$$
\begin{aligned}
\mathcal{E}_j&=\left|\tilde{\mathcal{F}}_j(\boldsymbol\theta)-\mathcal{F}_j(\boldsymbol\theta)\right|
=\left|\frac{\langle\mathcal{A}\rangle\pm\epsilon_{\mathcal{A}}}{\langle\mathcal{B}\rangle\pm\epsilon_{\mathcal{B}}}
-\frac{\langle\mathcal{A}\rangle}{\langle\mathcal{B}\rangle}\pm\epsilon_{\mathcal{O}}\right|\\
&\leq\eta_{1}^{-1}(\epsilon_{\mathcal{A}}+|\lambda_r|\epsilon_{\mathcal{B}})+\epsilon_{\mathcal{O}}.
\end{aligned}$$

\subsection{Sample cost}
Assuming that $M_k$($M_l$) is the number of samples used for measuring $\langle\mathcal{A}_k\rangle$($\langle\mathcal{B}_l\rangle$), the precision of each term is given by
\begin{equation}
\epsilon_a^2=\frac{\alpha_a^2\sigma_a}{M_a},\epsilon_b^2=\frac{\alpha_b^2\sigma_b}{M_b},
\end{equation}
where $\sigma_a(\sigma_b)$ represents the variance of the expectation value of $\mathcal{A}_a(\mathcal{B}_b)$. The total error of $\langle\mathcal{A}\rangle$, $\langle\mathcal{B}\rangle$ and the overlap $\sum_{i=1}^{j-1}\gamma_i|\langle\psi(\boldsymbol\theta)|\mathcal{B}|\psi(\boldsymbol\theta_{i}^{*})\rangle|^2$ scales as
\begin{equation}
\begin{aligned}
&\epsilon_\mathcal{A}^2=\sum_{k=0}^{K-1}\frac{\alpha_k^2\sigma_k}{M_k},
\epsilon_\mathcal{B}^2=\sum_{l=0}^{L-1}\frac{\alpha_l^2\sigma_l}{M_l},
\epsilon_O^2=\sum_{i=1}^{j-1}\frac{\gamma_i^2\sigma_i}{M_{i}},
\end{aligned}
\end{equation}
where $M_{i}$ and $\sigma_i$ are the number of samples and variance used for measuring $|\langle\psi(\boldsymbol\theta)|\mathcal{B}|\psi(\boldsymbol\theta_{i}^{*})\rangle|^2$. To determine the total sample complexity of finding the $j^{\textrm{th}}$ generalized eigenvalue $M^j=\sum_{k=0}^{K-1}M_k+\sum_{l=0}^{L-1}M_l+\sum_{i=1}^{j-1}M_i$, we define a quantity $\epsilon^2$ called pseudo-error which has a form
\begin{equation}\label{variance}
\begin{aligned}
\epsilon^2&=\epsilon_\mathcal{A}^2+\epsilon_\mathcal{B}^2+\epsilon_O^2.
\end{aligned}
\end{equation}
The real question now is how to choose the optimal choice of $M_k$, $M_l$ and $M_i$ by minimizing $\epsilon^2$ for the fewest measurements. We start with the Lagrange function
$$
\mathcal{L}=M^{j}+\mu
\bigg(\sum_{k=0}^{K-1}\frac{\alpha_k^2\sigma_k}{M_k}+\sum_{l=0}^{L-1}\frac{\alpha_l^2\sigma_l}{M_l}
+\sum_{i=1}^{j-1}\frac{\gamma_i^2\sigma_i}{M_{i}}-\epsilon^2\bigg).
$$
Our goal is to solve the following expression for $M_k$, $M_l$ and $M_i$,
\begin{equation}
\begin{aligned}
\min_{M_k,M_l,M_i}\max_{\mu}\mathcal{L}=\min_{M_k,M_l,M_i}M^j.
\end{aligned}
\end{equation}
We take the derivative of $\mathcal{L}$ with respect to $M_k,M_l,M_i$ to find,
$$
\begin{aligned}
&\frac{\partial\mathcal{L}}{\partial M_k}=\sum_{k=0}^{K-1}\Big(1-\mu\frac{\alpha_k^2\sigma_k}{M_k^2}\Big)=0
\rightarrow M_k=\sqrt{\mu\sigma_k}\alpha_k\sigma_k,\\
&\frac{\partial\mathcal{L}}{\partial M_l}=\sum_{l=0}^{L-1}\Big(1-\mu\sum_{l=0}^{L-1}\frac{\beta_l^2\sigma_l^2}{M_l}\Big)=0
\rightarrow M_l=\sqrt{\mu\sigma_l}\beta_l,\\
&\frac{\partial\mathcal{L}}{\partial M_i}=\sum_{i=1}^{j-1}\Big(1-\mu\frac{\gamma_i^2\sigma_i^2}{M_i^2}\Big)=0
\rightarrow M_i=\sqrt{\mu\sigma_i}\gamma_i.
\end{aligned}
$$
Taking this back into Eq. (\ref{variance}), we find that
\begin{align}
\sqrt{\mu}&=\frac{1}{\sigma^2}\Big(\sum_{k=0}^{K-1}\alpha_k\sqrt{\sigma_k}
+\sum_{l=0}^{L-1}\beta_l\sqrt{\sigma_l}+\sum_{i=1}^{j-1}\gamma_i\sqrt{\sigma_i}\Big)\nonumber\\
&=\frac{1}{\sigma^2}(\Lambda_k+\Lambda_l+\Lambda_i).
\end{align}
Now we can then learn
$$
\begin{aligned}
&M_k=\frac{\Lambda_k+\Lambda_l+\Lambda_i}{\epsilon^2}\alpha_k\sqrt{\sigma_k},\\
&M_l=\frac{\Lambda_k+\Lambda_l+\Lambda_i}{\epsilon^2}\beta_l\sqrt{\sigma_l},\\
&M_i=\frac{\Lambda_k+\Lambda_l+\Lambda_i}{\epsilon^2}\gamma_i\sqrt{\sigma_i}.
\end{aligned}
$$
The total number of measurement required is
\begin{align}
M^{j}&=\sum_{k=0}^{K-1}M_k+\sum_{l=0}^{L-1}M_l+\sum_{i=1}^{j-1}M_i\nonumber\\
&=\frac{1}{\epsilon^2}(\Lambda_k+\Lambda_l+\Lambda_i)^2.
\end{align}
Thus the total sample complexity of finding the $j^{\textrm{th}}$ nearly is
$$
\begin{aligned}
M&=M^{0}+M^{1}+\cdots+M^{j}\\
&=\frac{1}{\epsilon^2}\sum_{j^{'}=0}^{j-1}(\Lambda_a+\Lambda_b+\Lambda_i)^2\\
&\leq\frac{1}{\epsilon^2}\sum_{j^{'}=0}^{j-1}\Bigg(\sum_{k=0}^{K-1}\alpha_k+\sum_{l=0}^{L-1}\beta_l+\sum_{i=0}^{j^{'}-1}\gamma_i\Bigg)^2\\
&\leq\frac{j}{\epsilon^2}\Bigg(\sum_{k=0}^{K-1}\alpha_k+\sum_{l=0}^{L-1}\beta_l+\lambda_{r}\Bigg)^2\\
&=O\Bigg(\frac{j\Lambda^2}{\epsilon^2}\Bigg)
\end{aligned}
$$
for $\Lambda=\sum_{k=0}^{K-1}\alpha_k+\sum_{l=0}^{L-1}\beta_l+\lambda_{r}$.
\section{Method to decompose any matrix}
In this section, we explain how to decompose a $n-$qubit matrix $A\in\mathbb{C}^{N\times N}$ into a linear combination of Pauli strings, such that
\begin{equation}
\begin{aligned}
A=\sum_{i=0}^{N-1}\bigotimes_{j=0}^{n-1}a^{j}A_{i}^{j},
\end{aligned}
\end{equation}
where $A_{i}^{j}\in\{I,X,Y,Z\}$ is a Pauli operation and $a^{j}$ is a real parameter. The optimal parameter is obtained by the optimizing the follow cost function
\begin{equation}
\begin{aligned}
\{a^{j}\}_{j=0}^{n-1}=\min_{a^j}\|A-\sum_{i=0}^{N-1}\bigotimes_{j=0}^{n-1}a^{j}A_{i}^{j}\|_{\textrm{HS}}^2,
\end{aligned}
\end{equation}
where $\|X\|_{\textrm{HS}}$ is the Hilbert-Schmidt norm. Note that this cost function is faithful, vanishing if and only if $A$ has a exact decomposition. Since the number of Pauli basis is $\mathcal{O}(4^{n})$, $A$ has at most $\mathcal{O}(4^{n})$ terms. However, under some special structure of $A$, the number of terms can be reduced to $\mathcal{O}(2n+1)$ \cite{liu2020variational}.

In our experiment, the Pauli basis of $2-$qubit system is
\begin{align}
\{&I\otimes I,I\otimes X,I\otimes Y,I\otimes Z,X\otimes I,X\otimes X,X\otimes Y,\nonumber\\&
X\otimes Z,Y\otimes I,Y\otimes X,Y\otimes Y,Y\otimes Z,Z\otimes I,Z\otimes X,\nonumber\\&
Z\otimes Y,Z\otimes Z\}.
\end{align}
Due to the fact that the matrix $\mathcal{A}$ has a form
\begin{equation}
\begin{aligned}
\mathcal{A}=
\begin{bmatrix}
1.8 & 0 & 0 & 0.2\\
0 & 1 & 0.2 & 0\\
0 & 0.2 & 1 & 0\\
0.2 & 0 & 0 & 0.2
\end{bmatrix},
\end{aligned}
\end{equation}
the Pauli basis under the structure of $\mathcal{A}$ can be reduced to only $8$ operators, such that
\begin{align}
\{&I\otimes I,I\otimes Z,X\otimes X,X\otimes Y,\nonumber\\
&Y\otimes X,Y\otimes Y,Z\otimes I,Z\otimes Z\}.\nonumber
\end{align}


\begin{thebibliography}{99}
\bibitem{Shor1994} P. Shor, In Proceedings 35th annual symposium on foundations of computer science (pp. 124–134). IEEE (1994).
\bibitem{Grover1996} L. K. Grover, \href{https://doi.org/10.1103/PhysRevLett.79.325}{Phys. Rev. Lett. \textbf{79}, 325 (1997)}.
\bibitem{HHL2009} A. W. Harrow, A. Hassidim, and S. Lloyd, \href{https://doi.org/10.1103/PhysRevLett.103.150502}{Phys. Rev. Lett. \textbf{103}, 150502 (2009)}.
\bibitem{QML2017} J. Biamonte, P. Wittek, N. Pancotti, P. Rebentrost, N. Wiebe, and S. Lloyd, \href{https://doi.org/10.1038/nature23474}{Nature \textbf{549}, 195 (2017)}.
\bibitem{liu2018quantum} N. Liu and P. Rebentrost, \href{https://doi.org/10.1103/PhysRevA.97.042315}{Phys. Rev. A \textbf{97}, 042315 (2018)}.
\bibitem{liang2019quantum} J.-M. Liang, S.-Q. Shen, M. Li, and L. Li, \href{https://doi.org/10.1103/PhysRevA.99.052310}{Phys. Rev. A \textbf{97}, 052310 (2019)}.
\bibitem{preskill2018quantum} J. Preskill, \href{https://doi.org/10.22331/q-2018-08-06-79}{Quantum \textbf{2}, 79 (2018)}.
\bibitem{peruzzo2014variational} A. Peruzzo, J. McClean, P. Shadbolt, M.-H. Yung, X.-Q. Zhou, P. J. Love, A. Aspuru-Guzik, and J. L. O'Brien, \href{https://doi.org/10.1038/ncomms5213}{Nat. Commun. \textbf{5}, 4213 (2014)}.
\bibitem{higgott2019variational} O. Higgott, D. Wang, and S. Brierley, \href{https://doi.org/10.22331/q-2019-07-01-156}{Quantum \textbf{3}, 156 (2019)}.
\bibitem{jones2019variational} T. Jones, S. Endo, S. McArdle, X. Yuan, and S. C. Benjamin, \href{https://doi.org/10.1103/PhysRevA.99.062304}{Phys. Rev. A \textbf{99}, 062304 (2019)}.
\bibitem{vogt2021preparing} N. Vogt, S. Zanker, J.-M. Reiner, M. Marthaler, T. Eckl, and A. Marusczyk, \href{https://doi.org/10.1088/2058-9565/abe568}{Quantum Sci. Technol. 6 035003 (2021)}.
\bibitem{li2017} Y. Li and S. C. Benjamin, \href{https://doi.org/10.1103/PhysRevX.7.021050}{Phys. Rev. X \textbf{7}, 021050 (2017)}.
\bibitem{mahdian2020incoherent} M. Mahdian and H. Davoodi Yeganeh, \href{https://doi.org/10.1007/s11128-020-02800-8}{Quantum Inf Process \textbf{19}, 285 (2020)}.
\bibitem{endo2020variational} S. Endo, J. Sun, Y. Li, S. C. Benjamin, and X. Yuan, \href{https://doi.org/10.1103/PhysRevLett.125.010501}{Phys. Rev. Lett. \textbf{125}, 010501 (2020)}.
\bibitem{benedetti2019parameterized} M. Benedetti, E. Lloyd, S. Sack, and M. Fiorentini, \href{https://doi.org/10.1088/2058-9565/ab4eb5}{Quantum Sci. Technol. \textbf{4}, 043001 (2019)}.
\bibitem{wang2020variational} X. Wang, Z. Song, and Y. Wang, \href{https://doi.org/10.22331/q-2021-06-29-483}{Quantum \textbf{5}, 483 (2021)}.
\bibitem{li2021optimizing} K. Li, S. Wei, P. Gao, F. Zhang, Z. Zhou, T. Xin, X. Wang, P. Rebentrost, and G. Long, \href{https://doi.org/10.1038/s41534-020-00351-5}{npj Quantum Inf \textbf{7}, 16 (2021)}.
\bibitem{larose2019variational} R. LaRose, A. Tikku, \'{E}. O'Neel-Judy, L. Cincio, and P. J.Coles, \href{https://doi.org/10.1038/s41534-019-0167-6}{npj Quantum Inform. \textbf{5}, 8 (2019)}.
\bibitem{nakanishi2019subspace} K. M. Nakanishi, K. Mitarai, and K. Fujii, \href{https://doi.org/10.1103/PhysRevResearch.1.033062}{Phys. Rev. Research \textbf{1}, 033062 (2019)}.
\bibitem{mcclean2017hybrid} J. R. McClean, M. E. Kimchi-Schwartz, J. Carter, and W. A. de Jong, \href{https://doi.org/10.1103/PhysRevA.95.042308}{Phys. Rev. A \textbf{95}, 042308 (2017)}.
\bibitem{parrish2019quantum} R. M. Parrish, E. G. Hohenstein, P. L. McMahon, and T. J. Mart\'{i}nez, \href{https://doi.org/10.1103/PhysRevLett.122.230401}{Phys. Rev. Lett. \textbf{122}, 230401 (2019)}.
\bibitem{wei2020a} S. Wei, H. Li, and G. Long, \href{https://doi.org/10.34133/2020/1486935}{Research \textbf{2020}, 1486935 (2020)}.
\bibitem{amos1969a} A. T. Amos, C. Laughlin, and G. R. Moody, \href{https://doi.org/10.1090/S0025-5718-1980-0583502-2}{Chem. Phys. Lett. \textbf{3}, 411 (1969)}.
\bibitem{cliffe2000the} K. A. Cliffe, A. Spence, and S. J. Tavener, \href{https://doi.org/10.1017/S0962492900000398}{Acta Numer. \textbf{9}, 39 (2000)}.
\bibitem{bittnar1996numerical} Z. Bittnar and J. \v{S}ejnoha, Oscillation Matrices and Kernels and Small Vibrations of Mechanical Systems. AMS, Providence (2002).
\bibitem{ghaboussi2016numerical} J. Ghaboussi and X. S. Wu, Numerical Methods in Computational Mechanics. CRC Press, Boca Raton (2016).
\bibitem{ford1974the} B. Ford and G. Hall, \href{https://doi.org/10.1016/0010-4655(74)90011-3}{Comput. Phys. Commun. \textbf{8}, 337 (1974)}.
\bibitem{gantmacher2002oscillation} F. P. Gantmacher and M. G. Krein, Oscillation Matrices and Kernels and Small Vibrations of Mechanical Systems. AMS, Providence (2002).
\bibitem{chugunova2010count} M. Chugunova and D. Pelinovsky, \href{https://doi.org/10.1063/1.3406252}{J. Math. Phys. \textbf{51}, 052901 (2010)}.
\bibitem{golub1989matriz} G. H. Golub and C. F. Van Loan, Matriz Computations. The Johns Hopkins University Press, Baltimore (1989).
\bibitem{ericsson1980the} T. Ericsson and A. Ruhe, \href{https://doi.org/10.1090/S0025-5718-1980-0583502-2}{Math. Comput. \textbf{35}, 1251 (1980)}.
\bibitem{sakuraia2003a} T. Sakuraia and H. Sugiura, \href{https://doi.org/10.1016/S0377-0427(03)00565-X}{J. Comput. Appl. Math. \textbf{159}, 119 (2003)}.
\bibitem{ikegamia2009a} T. Ikegamia, T. Sakuraib, and U. Nagashimac, \href{https://doi.org/10.1016/j.cam.2009.09.029}{J. Comput. Appl. Math. \textbf{233}, 1927 (2010)}.
\bibitem{parker2020quantum} J. B. Parker and I. Joseph, \href{https://doi.org/10.1103/PhysRevA.102.022422}{Phys. Rev. A \textbf{102}, 022422  (2020)}.
\bibitem{liu2020variational} H. Liu, Y. Wu, L. Wan, S. Pan, S. Qin, F. Gao, and Q. Wen, \href{https://doi.org/10.1103/PhysRevA.104.022418}{Phys. Rev. A \textbf{104}, 022418 (2021)}.
\bibitem{parlett1998symmetric} B. N. Parlett, The symmetric eigenvalue problem, vol. 20 (siam, 1998).
\bibitem{aharonov2008a} D. Aharonov, V. Jones, and Z. Landau, \href{https://doi.org/10.1007/s00453-008-9168-0}{Algorithmica \textbf{55}, 395 (2009)}.
\bibitem{nielsen2000quantum} M. A. Nielsen and I. L. Chuang, Quantum Computation and Quantum Information. Cambridge University Press, Cambridge (2000).
\bibitem{romero2018strategies} J. Romero, R. Babbush, J. R. McClean, C. Hempel, P. J. Love, and A. Aspuru-Guzik, \href{https://doi.org/10.1088/2058-9565/aad3e4}{Quantum Sci. Technol. \textbf{4}, 014008 (2019)}.
\bibitem{buhrman2001quantum} H. Buhrman, R. Cleve, J. Watrous, and R. de Wolf, \href{https://doi.org/10.1103/PhysRevLett.87.167902}{Phys. Rev. Lett. \textbf{87}, 167902 (2001)}.
\bibitem{carlos2013swap} J. C. Garcia-Escartin and P. Chamorro-Posada, \href{https://doi.org/10.1103/PhysRevA.87.052330}{Phys. Rev. A \textbf{87}, 052330 (2013)}.
\bibitem{cincio2018learning} L. Cincio, Y. Subasi, A. T. Sornborger, and P. J. Coles, \href{https://doi.org/10.1088/1367-2630/aae94a}{New J. Phys. \textbf{20}, 113022 (2018)}.
\bibitem{havlivcek2019supervised} Vojt\v{e}ch Havl\'{i}\v{c}ek, A. D. C\'{o}rcoles, K. Temme, A. W. Harrow, A. Kandala, J. M. Chow, and J. M. Gambetta, \href{https://doi.org/10.1038/s41586-019-0980-2}{Nature \textbf{567}, 209 (2019)}.
\bibitem{kingma2014adma} D. P. Kingma and J. Ba, \href{http://arxiv.org/abs/1412.6980}{arXiv:1412.6980}.
\bibitem{wierichs2020avoiding} D. Wierichs, C. Gogolin, and M. Kastoryano, \href{https://doi.org/10.1103/PhysRevResearch.2.043246}{Phys. Rev. Res. \textbf{2}, 043246 (2020)}.
\bibitem{dean2020avoiding} J. Rivera-Dean, P. Huembeli, A. Ac\'{i}n, J. Bowles, \href{https://export.arxiv.org/abs/2104.02955}{arXiv:2104.02955}.
\bibitem{HardwareVQE2017} A. Kandala, A. Mezzacapo, K. Temme, M. Takita, M. Brink, J. M. Chow, and J. M. Gambetta, \href{https://doi.org/10.1038/nature23879}{Nature \textbf{549}, 242 (2017)}.
\bibitem{fisher1936the} R. A. Fisher, \href{https://doi.org/10.1111/j.1469-1809.1936.tb02137.x}{Ann. eugen. \textbf{7}, 179 (1936)}.
\bibitem{abrams1999quantum} D. S. Abrams and S. Lloyd, \href{https://doi.org/10.1103/PhysRevLett.83.5162}{Phys. Rev. Lett. \textbf{83}, 5162 (1999)}.
\bibitem{jones2012faster} N. C. Jones, J. D. Whitfield, P. L. McMahon, M.-H. Yung, R. V. Meter, A. Aspuru-Guzik and Y. Yamamoto, \href{https://doi.org/10.1088/1367-2630/14/11/115023}{New J. Phys. \textbf{14} 115023 (2012)}.
\bibitem{liang2019variational} J.-M. Liang, S.-Q. Shen, M. Li, and L. Li, \href{https://doi.org/10.1103/PhysRevA.101.032323}{Phys. Rev. A \textbf{101}, 032323 (2020)}.
\bibitem{long2006general} G.-L. Long, \href{https://doi.org/10.1088/0253-6102/45/5/013}{Commun. Theor. Phys. \textbf{45}, 825 (2006)}.
\bibitem{berry2015simulating} D. W. Berry, A. M. Childs, R. Cleve, R. Kothari, and R. D. Somma, \href{https://doi.org/10.1103/PhysRevLett.114.090502}{Phys. Rev. Lett. \textbf{114}, 090502 (2015)}.
\bibitem{arbenz2016solving} P. Arbenz, Tech. Rep., Computer Science Department, ETH Z\"{u}rich (2016).
\bibitem{golub2002an} G. H. Golub and Q. Ye, \href{https://doi.org/10.1137/S1064827500382579}{SIAM J. Sci. Comput. \textbf{24}, 312 (2002)}.
\bibitem{long2001efficient} G.-L. Long and Y. Sun, \href{https://doi.org/10.1103/PhysRevA.64.014303}{Phys. Rev. A \textbf{64}, 014303 (2001)}.
\bibitem{vazquez2001efficient} A. C. Vazquez and S. Woerner, \href{https://doi.org/10.1103/PhysRevApplied.15.034027}{Phys. Rev. Applied \textbf{15}, 034027 (2021)}.
\bibitem{soklakov2006quantum} A. N. Soklakov, R. Schack, \href{https://arxiv.org/abs/quant-ph/0408045}{arXiv:quant-ph/0408045}.
\bibitem{giovannetti2008quantum} V. Giovannetti, S. Lloyd, and L. Maccone, \href{https://doi.org/10.1103/PhysRevLett.100.160501}{Phys.Rev.Lett. \textbf{100}, 160501 (2008)}.
\bibitem{barenco1995elementary} A. Barenco, C. H. Bennett, R. Cleve, D. P. DiVincenzo, N. Margolus, P. Shor, T. Sleator, J. A. Smolin, and H. Weinfurter, \href{https://doi.org/10.1103/PhysRevA.52.3457}{Phys. Rev. A \textbf{52} 3457 (1995)}.
\bibitem{paddlequantum} \href{https://github.com/paddlepaddle/Quantum}{Paddle Quantum, (2020).}
\bibitem{paddle2} Y. Ma, D. Yu, T. Wu, and H. Wang, \href{http://www.jfdc.cnic.cn/EN/abstract/abstract2.shtml}{Frontiers of Data and Domputing \textbf{1}, 105 (2019)}.
\bibitem{bishop2006pattern} C. M. Bishop, Pattern Recognition and Machine Learning. Springer Science, Heidelberg (2006).
\bibitem{huggins2020a} W. J. Huggins, J. Lee, U. Baek, B. O'Gorman, and K. B. Whaley, \href{https://doi.org/10.1088/1367-2630/ab867b}{New J. Phys. \textbf{22} 073009 (2020)}.
\bibitem{sakurai2010a} T. Sakurai, H. Tadano, T. Ikegami, and U. Nagashima, \href{https://doi.org/10.11650/twjm/1500405871}{Taiwan. J. Math. \textbf{14}, 855 (2010)}.
\bibitem{teng2011efficient} H. Teng, T. Fujiwara, T. Hoshi, T. Sogabe, S. L. Zhang, and S. Yamamoto, \href{https://doi.org/10.1103/PhysRevB.83.165103}{Phys. Rev. B \textbf{83}, 165103 (2011)}.
\bibitem{woitzik2020entanglement} A. J. C. Woitzik, P. K. Barkoutsos, F. Wudarski, A. Buchleitner, and I. Tavernelli, \href{https://doi.org/10.1103/PhysRevA.102.042402}{Phys. Rev. A \textbf{102}, 042402 (2020)}.
\bibitem{cerezo2020variational} M. Cerezo, K. Sharma, A. Arrasmith, and P. J. Coles, \href{https://arxiv.org/abs/2004.01372}{arXiv:2004.01372}.
\bibitem{mcclean2018barren} J. R. McClean, S. Boixo, V. N. Smelyanskiy, R. Babbush, and H. Neven, \href{https://www.nature.com/articles/s41467-018-07090-4}{Nat. Commun. \textbf{9}, 4812 (2018)}.
\end{thebibliography}
\end{document}